\documentclass[aip,jcp,reprint]{revtex4-1}
\usepackage[utf8]{inputenc}
\usepackage{amsmath} 
\usepackage{amsfonts}
\usepackage{mathrsfs}
\usepackage{bm}
\usepackage{amssymb}
\usepackage{graphicx}  
\usepackage{braket}
\usepackage{grffile}
\usepackage{dsfont}
\usepackage{comment}
\usepackage{hyperref}
\usepackage{dcolumn}

\newcommand{\pd}[2]{\frac{\partial{#1}}{\partial{#2}}}

\newcommand{\mk}[1]{\overline{#1}}

\newcommand{\KK}[0]{\mathcal{K}}
\newcommand{\LL}[0]{\mathcal{L}}
\newcommand{\I}[0]{\mathcal{I}}
\newcommand{\GG}[0]{\mathcal{G}}

\newcommand{\PP}[0]{\mathcal{P}}
\newcommand{\QQ}[0]{\mathcal{Q}}

\newcommand{\WW}[0]{\mathcal{W}}
\newcommand{\TTp}[0]{\mathcal{T}_+}

\newcommand{\UU}[0]{\mathcal{U}}
\newcommand{\dd}[2]{\frac{d{#1}}{d{#2}}}

\newcommand{\tr}[0]{\text{tr}}
\newcommand{\proj}[2]{| #1 \rangle \! \langle #2 |}
\newcommand{\pproj}[2]{| #1 \rangle\!\rangle \! \langle\!\langle #2 |}
\newcommand{\bbra}[1]{\langle\!\langle{#1}|}
\newcommand{\kket}[1]{|{#1}\rangle\!\rangle}
\newcommand{\bbrakket}[1]{\langle\!\langle{#1}\rangle\!\rangle}
\newcommand{\trb}[0]{\tr_{\rm ph}}
\newcommand{\RE}[0]{\,{\rm Re}\,}

\newcommand{\Uv}[2]{U_{#1}^{#2}(t)}

\begin{document}

\title{Excitation energy transfer rates: comparison of approximate methods to the exact solution}

\author{Simon Jesenko}
 \email{simon.jesenko@fmf.uni-lj.si}
\author{Marko \v Znidari\v c}%

\affiliation{
Faculty of Mathematics and Physics, University of Ljubljana
}%

\date{\today}

\begin{abstract}
We have evaluated excitation energy transfer rates in photosyntetic complexes using the exact HEOM method and various approximate methods frequently used in the literature, namely, the F\"orster method, the Redfield method, the modified Redfield method and the variational master equation. The rates are evaluated for the case of a simple dimer and a trimer photosynthetic complex, with vibrational environment characterized by the Drude-Lorentz spectral density. Comparing approximate rates to the exact ones, we have confirmed the validity of approximate methods within appropriate limits, however these limits are often well outside parameter ranges that are relevant for the dynamics in real photosynthetic complexes.
\end{abstract}

\pacs{87.15.M-, 87.14.E-, 87.15.H-, 82.50.Hp, 33.80.-b, 05.30.-d, 02.50.Ga}
\maketitle

\section{Introduction}

In recent years, the role of quantum mechanics in the excitation energy transfer (EET) in photosynthesis has been widely debated, following the observations of long-lived oscillations in the two-dimensional (2D) spectra in various photosynthetic complexes (PPCs)\cite{Lee2007,Engel2007,Collini2010, Panitchayangkoon2010}. Possible mechanisms that could be employed in nature to improve the efficiency of the process were investigated\cite{Rebentrost2009a, Ishizaki2010PCCP, Pachon2011, Ishizaki2009,Plenio2008, Rebentrost2009,Mohseni2008,Caruso2009,Wu2010}. However, the relation between the experimentally measured dynamics of the PPC and the \textit{in vivo} dynamics due to incoherent light illumination was not clear\cite{Brumer2012,Kassal2013,Mancal2010,Cheng2009,Fassioli2012,Tiersch2012}. In our recent work\cite{Jesenko2013} we have shown that the efficiency of the EET process \textit{in vivo} is completely determined by the \textit{rate} kernel $\mk K$, being equal to the time-integral of the memory kernel $K(t)$ of population dynamics of electronic excitations. The rate kernel $\mk K$ already includes all contributions due to quantum mechanical nature of dynamics. Therefore it can be conveniently used for the comparison of various approximations. Evaluation of rate kernel $\mk K$ from the underlying microscopic model is however far from trivial. In this paper we are considering different approaches for its evaluation, comparing exact results obtained by the HEOM method to the various perturbative approaches.

The rate kernel is difficult to evaluate due to a specific strength of the environmental interaction, i.e., coupling between electronic and vibrational degrees of freedom (DOFs), which is of comparable magnitude as other energy scales in the system. Thus one can not reliably employ perturbation theory with respect to environmental interaction, nor with respect to the inter-pigment coupling. At the same time, due to large number of environmental DOFs, the exact treatment of all vibrational modes is usually not feasible -- the dimensionality of the corresponding Hilbert space grows exponentially with the number of normal modes considered. For small number of pigment molecules and specific forms of environmental interaction, we can nonetheless calculate dynamics exactly using various numerical methods. Examples of such methods include the quasi-adiabatic propagator path integral (QUAPI)~\cite{Makri1995a,Makri1995b,Nalbach2011}, the multiconfiguration time-dependent Hartree (MCTDH)~\cite{Meyer1990, Beck2000,Seibt2009}, and the hierarchical equations of motion (HEOM)~\cite{Tanimura1989,Ishizaki2009a,Shi2009}. We will adapt the HEOM method for the evaluation of the kernel $\mk K$. The analysis will be limited to the Drude-Lorentz spectral density as the HEOM method is less efficient for other spectral densities.

Exact methods are computationally very demanding and also rather involved to implement. Also, they usually do not provide simple and intuitive picture of processes involved in the EET. Thus, various approximate descriptions of EET dynamics were introduced, that are believed to encapsulate the main characteristics of the exact dynamics. They are based on the perturbation theory, where one tries to separate Hamiltonian to the exactly solvable part and small interaction part, which can then be treated as a perturbative contribution to the exact dynamics. We will derive expressions for the rate kernel $\mk K$ within these approximations. We will focus on methods that are frequently used in studies of EET dynamics, namely the \textit{F\"orster theory}~\cite{forster1948}, the \textit{Redfield theory}~\cite{redfield1957theory}, the \textit{modified Redfield theory}~\cite{Zhang1998} and the \textit{variational master equation}~\cite{Yarkony1977, Silbey1984,McCutcheon2011,Zimanyi2012}. A unified derivation of kernels for all these methods based on a projection operator formalism will be presented. Where possible, we will formulate a method by deriving the quantum Markovian kernel $\mk \KK$, which can then be mapped to the classical rate kernel $\mk K$ in an arbitrary basis. Some of the approximate methods are however from the very beginning limited to the description of dynamics in terms of populations via classical rate kernel $\mk K$ in a specific basis. Results from approximate methods will be compared to the exact HEOM calculations, providing a critical assessment of their applicability in the context of EET dynamics.

Some aspects of approximate methods were already analyzed in the literature. EET rates for the F\"orster, Redfield and modified Redfield theory were analyzed in the work of Yang and Flemming\cite{Yang2002}, however, they were not compared to the rates obtained from an exact calculation. These three approximate approaches were also compared in Ref.~\onlinecite{Novoderezhkin2013} for the case of a B800 ring in the LH2 antenna complex, focusing on the absorption spectra and population dynamics. Generalization of the F\"orster theory, intended for a description of weakly coupled clusters of PPCs, was also considered in the analysis in Ref~\onlinecite{Novoderezhkin2010}. Applicability of approximations was also discussed by Ishizaki et al.\cite{Ishizaki2009} Various approximate methods were compared in terms of resulting absorption spectra in Refs.~\onlinecite{Schroder2006,Schroder2007}. For the variational master equation, approximate density matrix dynamics was compared to the exact one for the \textit{super-Ohmic} spectral density in Ref.~\onlinecite{Lee2012}. Our analysis complements the existing results by considering the approximations on the basis of the rate kernel, comparing the approximate rates to the exact ones.

The outline of the paper is as follows. In section~\ref{sec:model} we first introduce the relevant microscopic model of the EET, and then review the mapping of full-system dynamics to the description at the level of electronic excitations, leading to the corresponding quantum kernel $\mk \KK$ and the rate kernel $\mk K$. In section~\ref{sec:methods} we will present the exact HEOM method and various approximate methods, all adapted for the calculation of the rate kernel $\mk K$. Then, in section~\ref{sec:comparison}, the exact and approximate kernels will be compared for a two site PPC (dimer) and a three site PPC (trimer). We will conclude in section~\ref{sec:conclusion} with a discussion of the applicability of various approximate methods in different parameter regimes.

\section{Model}
\label{sec:model}
\subsection{Microscopic model}
\label{sec:microscopic_model}

The main constituents of photosynthetic complexes are pigment molecules with electronic transitions in the range of visible light, and solvent molecules (proteins, water molecules, etc.), which provide structure to the PPC. Note that we also consider proteins as the solvents, although they can be of comparable size or even larger than the pigment molecules. The microscopic state of the PPC is specified by a set of electronic and nuclear coordinates of all constituents, and the dynamics of the system is determined by the corresponding molecular Hamiltonian, accounting for their kinetic energy and Coulomb interaction among them. Employing the usual approximations~\cite{may2011charge}, we end up with the Hamiltonian
\begin{equation}
H = H_{\rm el} + H_{\rm el-ph} + H_{\rm ph},
\label{eq:H_ppc}
\end{equation}
which is comprised of electronic, phonon and interaction part,
\begin{align}
  \label{eq:H_el}
  H_{\rm el} &= \sum_{m=1}^N E_m \proj{m}{m} + \frac{1}{2} \sum_{m,n = 1}^N V_{mn} \proj{m}{n},\\
  \label{eq:H_ph}
  H_{\rm ph} &= \sum_{m=1}^N \sum_\xi \omega_{m\xi} b_{m\xi }^\dagger b_{m\xi},\\
  \label{eq:H_el-ph}
  H_{\rm el-ph} &= \sum_{m=1}^N \sum_\xi g_{m\xi} (b_{m\xi}^\dagger + b_{m\xi}) \proj{m}{m}.
\end{align}
The summation goes over $N$ pigment molecules in a PPC. Vector denoted by $\ket{m}$ corresponds to the state in which the $m$th pigment molecule is excited, while the others are in the ground state, namely $\ket{m} =  \ket{\varphi_{me}} \prod_{n\neq m} \ket{\varphi_{ng}}$, with $\ket{\varphi_{me}}$ denoting the first excited electronic state of the pigment molecule, and $\ket{\varphi_{mg}}$ the ground electronic state of the pigment molecule. We can limit the discussion to this single-excited subspace due to low sunlight intensity, leading to negligible probability of more than one pigment molecule being excited simultaneously. The basis spanned by $\{ \ket{m} \}$ is known as the \textit{site} basis. Site energy $E_m$ corresponds to the energy difference between the ground and the excited electronic state of the pigment molecule in the absence of the interaction term $H_{\rm el-ph}$. Exciton coupling $V_{mn}$ is governed by the transition dipole moments for the transition from the ground to the first excited state of given molecules.

The vibrational Hamiltonian $H_{\rm ph}$ is given by a set of vibrational modes enumerated by $\xi$, with frequencies $\omega_{m \xi}$, while $b^\dagger_{m \xi}$ in $b_{m \xi}$ are usual creation/annihilation bosonic operators. Coupling between electronic and vibrational DOFs is determined by parameters $g_{m\xi}$. The strength of this coupling can be characterized by a \textit{reorganization energy} $\lambda_m = \sum_\xi g_{m\xi}^2/\omega_{m\xi}$, which largely determines the properties of EET dynamics. The reorganization energy is equal to the difference between equilibrium energy of vibrational modes in the ground state and the excited state of the pigment molecule.

Summation over individual vibrational modes can be replaced by frequency-integration via introduction of a \textit{spectral density}
\begin{equation}
J_m(\omega) = \sum_\xi g_{m \xi}^2 \delta(\omega-\omega_{m\xi}),
\label{eq:J_m}
\end{equation}
where individual vibrational mode correspond to Dirac delta function at frequency $\omega_{m\xi}$. If the environment consists of a large number of closely-spaced vibrational modes, the spectral function can be approximated by a continuous function. We limit the analysis to the \textit{Drude-Lorentz} spectral density,
\begin{equation}
  J_m(\omega) = \frac{2}{\pi} \lambda_m \frac{\gamma_m \omega}{\omega^2 + \gamma_m^2},
  \label{eq:drude-lorentz}
\end{equation}
where $\lambda_m$ is the reorganization energy defined above, and $\gamma_m$ determines the cut-off frequency for the vibrational modes.

\subsection{Equations of motion}
\label{sec:EOM}
Dynamics of the whole PPC is governed by the Hamiltonian $H$ from Eq.~\eqref{eq:H_ppc} via quantum Liouville equation
\begin{equation}
\dd{R(t)}{t}=-i[H ,R(t)] \equiv \LL R(t),
\label{eq:eom_whole_system}
\end{equation}
where $R(t)$ is the density matrix for all DOFs, including electronic and vibrational part, while $\LL$ is the full-system Liouvillian. If we are interested only in the state of electronic DOFs, we can obtain the electronic density matrix by taking a partial trace over vibrational DOFs, $\rho(t) = \trb \, R(t)$. However, it is more insightful if we describe the dynamics of $\rho(t)$ directly, without referring to the vibrational DOFs in $R(t)$ -- this is specially true when one is considering the efficiency of EET process, for details see also Ref.~\onlinecite{Jesenko2013}. Equations of motion on a given subspace can be obtained via the projection operator technique (i.e., the Nakajima-Zwanzig formalism)\cite{breuer2002theory}. Such equations reproduce dynamics within a given subspace exactly. We can employ the projection operator formalism to either obtain the description on the level of electronic density matrix $\rho(t)$, or even at the level of populations, i.e. diagonal elements of the density matrix $\rho$ in a certain basis. In the following we present both mappings.

We define projection operator to the electronic density matrix $\PP$ via relation
\begin{equation}
 \PP R = (\trb R) \otimes \rho_{\rm ph}.
 \label{eq:PP_def}
\end{equation}
where $\rho_{\rm ph}$ is some reference state of the phonon environment. Projection operator formalism enables us to write a dynamical equation for $\rho(t)$ in the form of a \textit{generalized quantum master equation},
\begin{equation}
  \dd{\rho(t)}{t}=\int_{0}^{t}ds \, \KK(s) \rho(t-s),
  \label{eq:generalized_master_eq}
\end{equation}
where the kernel $\KK(t)$ is given by
\begin{equation}
\KK(t) = \PP \LL \PP \delta(t) + \PP \LL \GG_{\QQ}(t) \QQ \LL \PP,
\label{eq:KKt}
\end{equation}
with the propagator $\GG_\QQ(t) = \exp\left[\QQ \LL t \right]$, where $\QQ$ is a projector to the irrelevant part, given by $\QQ = 1 - \PP$. When describing dynamics via a generalized master equation, the effect of vibrational DOFs on the dynamics of electronic excitations is encoded in time-dependence of kernel $\KK(t)$. Note that description with the generalized master equation \eqref{eq:generalized_master_eq} is valid only when the initial condition for the whole system is chosen such that $\QQ R(0) = 0$, e.g., product state of electronic and vibrational density matrices $R(0)=\rho(0) \otimes \rho_{\rm ph}$. Otherwise, the explicit dependence on the irrelevant part of the initial condition $\QQ R(0)$ must also be taken into account.

When the density matrix $\rho(t)$ is changing at time-scales much longer than the time-span of the time-dependent kernel $\KK(t)$, the Markovian approximation can be employed, which treats $\rho(t)$ as being constant for the duration of the kernel. This results in a \textit{quantum master equation},
\begin{equation}
  \dd{\rho(t)}{t} = \mk\KK \rho(t).
  \label{eq:master_eq}
\end{equation}
where the Markovian kernel $\mk \KK$ is obtained as $\mk\KK = \int_0^\infty dt\, \KK(t)$. Even when the Markovian approximation is not justified on the basis of time-scale separation, the given Markovian kernel is still relevant for the analysis of EET, as it exactly determines the stationary currents and the efficiency of the process (see Ref.~\onlinecite{Jesenko2013} for details). Thus we will be comparing different approximations with respect to the resulting Markovian kernels $\mk \KK$.

Analogous projection operator formalism as above can be also employed to obtain dynamics of populations only, i.e. the dynamics of  diagonal elements of density matrix $\rho(t)$ in some basis, $\bm p = (\rho_{11},\dots,\rho_{NN})$. In this case, we employ the projection operator
\begin{equation}
  \PP_c R = \sum_{m} \proj{m}{m} \, \trb \braket{m|R|m} \otimes \rho_{\rm ph}^{m},
  \label{eq:PP_c}
\end{equation}
where $\rho_{\rm ph}^{m}$ are some reference states of the environment. We have included possible site-dependence of the reference environmental states $\rho_{\rm ph}^m$ to enable derivation of F\"orster and related approaches from this formalism in the following sections. Projector to the irrelevant part is again defined as $\QQ_c = \I - \PP_c$. The corresponding projection operator formalism results in a \textit{generalized classical master equation}
\begin{equation}
  \dd{\bm p(t)}{t}=\int_{0}^{t} ds \,K(t-s) \bm p(s),
  \label{eq:generalized_master_eq_classical}
\end{equation}
which reproduces dynamics of populations exactly (for the appropriately chosen initial condition $R(0)$). The kernel $K(t)$ is given by the expression analogous to Eq.~\eqref{eq:KKt}, replacing projectors $\PP$ and $\QQ$ with $\PP_c$ and $\QQ_c$. The term ``classical'' only denotes the fact that the equation does not include off-diagonal elements of the density matrix, while the quantum nature of the underlying model is still completely accounted for via the time-dependence of the kernel $K(t)$.

Employing the Markovian approximation, a \textit{classical master equation} is obtained,
\begin{equation}
  \dd{\bm p(t)}{t}= \mk K \bm p(t),
  \label{eq:master_eq_classical}
\end{equation}
where the Markovian kernel is given by $\mk K = \int_0^\infty dt\, K(t)$. We will call $\mk K$ the \textit{rate} kernel, as its individual entries $\mk K_{nm}$ can be interpreted as probability rates for the transfer of excitation from the $m$th to the $n$th pigment. Similarly as in the case of quantum master equation, the Markovian kernel $\mk K$ is of physical relevance also when the Markovian approximation does not lead to appropriate short-time dynamics -- it nonetheless completely determines the stationary currents between populations and with it the efficiency of the EET process.

\section{Methods}
\label{sec:methods}

In this section, we present the methods that will be used in the calculation of the rate kernel $\mk K$. Where possible, we will first derive the expression for the full Markovian quantum kernel $\mk \KK$, from which the corresponding rate kernel $\mk K$ in an arbitrary basis can be obtained from the Nakajima-Zwanzig equation for the projection to the diagonal elements,
\begin{equation}
\PP_{p} \rho = \sum_m \proj{m}{m} \braket{m|\rho|m}.
\label{eq:PPp}
\end{equation}
The exact HEOM method will be adapted for the calculation of the exact kernel $\mk \KK$ by employing projection operator formalism on the hierarchy of equations which determines the exact evolution of electronic density matrix $\rho(t)$.

All approximate approaches that will be compared to the HEOM method are based on perturbation theory, where the complete Hamiltonian $H$ from Eq.~\eqref{eq:H_ppc} is decomposed to an exactly solvable part $H_0$ and a small perturbative part $H_I$. Depending on this decomposition, approximative approaches are valid in different parameter regimes. An approximate Markovian quantum kernel can be obtained from the exact expression for $\KK(t)$, Eq.~\eqref{eq:KKt}, by a second order expansion of $H_I$ contribution, resulting in the approximate kernel
\begin{equation}
\mk \KK \approx \PP \LL_0 \PP + \int_0^\infty dt \, \PP \LL_I \exp{(\LL_0 t)} \LL_I \PP,
\label{eq:mkKK_second_order}
\end{equation}
where Liouvillians $\LL_0$ a $\LL_I$ correspond to $H_0$ in $H_I$. The reference state of phonons in $\PP$, Eq.~\eqref{eq:PP_def}, is given by $\rho_{\rm ph}=\tr_{\rm el}\{e^{-\beta H_0}\}/Z$, where we have denoted a partial trace over electronic subspace and $Z = \tr \{e^{-\beta H_0}\}$. This expression will be employed in the case of the Redfield theory and the variational master equation.

For other two approximate methods, namely the F\"orster theory and the modified Redfield theory, the reference state of the phonon environment is chosen to be site-dependent. We are thus limited to the description in terms of the rate kernel $\mk K$, as projection operator $\PP_c$ from Eq.~\eqref{eq:PP_c} can account for the site-dependence of phonon reference states. In this case a second-order expansion of $H_I$ in the expression for $K(t)$ results in equation for individual elements of the rate kernel\cite{Yang2002}
\begin{equation}
\mk K_{ab} \approx 2 \RE \int_0^\infty dt\, \trb \{ e^{i {H_0^b} t} \braket{b|H_I|a}e^{-i {H_0^a} t} \braket{a|H_I|b} \rho_{\rm ph}^{b} \},
\label{eq:mkK_second_order_explicit}
\end{equation}
where $H_0^a = \braket{a|H_0|a}$ and $\rho_{\rm ph}^a=e^{-\beta H_{\rm 0}^a}/Z$, while $\{\ket{a} \}$ are some basis vectors spanning the electronic subspace.

\subsection{Exact HEOM method}

The exact HEOM (hierarchical equations of motion) method was initially used in the context of EET for the calculation of time-dependence of density matrix $\rho(t)$.\cite{Ishizaki2009a} Here we adapt it for the evaluation of the time-dependent kernel $\KK(t)$ and the Markovian kernel $\mk \KK$. In the HEOM method the effect of vibrational DOFs on the dynamics of electronic excitations is encoded in auxiliary operators, the dynamics of which is governed by a system of linear differential equations. The system of differential equations has a convenient hierarchical structure, which is practical for numerical evaluation. The usual derivation of the HEOM method is based on the the Feynman path integral formalism~\cite{Tanimura1989,Ishizaki2005,Xu2005,Xu2007}, or, alternatively, on the stochastic approach~\cite{Yan2004,Zhou2008,Schroder2007b}.

The exact solution for the electronic density matrix in interaction picture, defined by the relation $\tilde \rho(t) = \exp(\LL_0 t) \rho(t)$ with $\LL_0 = \LL_{\rm el} + \LL_{\rm ph}$, can be formally written as $\tilde \rho(t) = \tilde{\UU}(t) \rho(0)$, where we have introduced propagator
\begin{align}
\label{eq:propagator_interaction_picture}
\tilde{\mathcal{U}}(t) &= \trb \left\{ \TTp \exp\left(\int_0^t ds\, \tilde \LL_{\rm el-ph}(s)\right) \rho_{\rm eq} \right\},
\end{align}
with $\rho_{\rm eq}=e^{-\beta H_{\rm ph}}/Z$ being equilibrium state of the phonon environment, and $\TTp$ is the usual \textit{time ordering operator}. Operators in interaction picture are given by $\tilde X(t) = \exp(-\LL_0 t) X$, and the superoperators are given by $\tilde{\mathcal{X}}(t) = \exp(-\LL_0 t) \mathcal{X} \exp(\LL_0 t)$. Assuming that each pigment is coupled only to its own independent set of vibrational DOFs and using the Wick theorem for harmonic oscillators\cite{Danielewicz1984} the propagator factorizes,
\begin{equation}
\tilde{\UU}(t) = \TTp \prod_{m=1}^N \exp\left(\int_0^t ds\,\tilde \WW_m(s) \right),
\label{eq:heom_propagator}
\end{equation}
where we have introduced
\begin{equation}
\begin{split}
\tilde{\WW}_m(s) = & - \int_0^s d\tau\, \tilde V_m(s)^\times \times  \\
& \left[C_m'(s-\tau) \tilde V_m(\tau)^\times + i C_m''(s-\tau) \tilde V_m(\tau)^\circ \right].
\end{split}
\label{eq:heom_WW}
\end{equation}
The functions $C_m'(t)$ and $C_m''(t)$ are the real and imaginary part of the correlation function of the electron-phonon interaction, $C_m(t) = \trb\left\{\tilde u_m(t) \tilde u_m(0) \rho_{\rm eq} \right\}$, with $\tilde u_m = \braket{m|\tilde H_{\rm el-ph}(t)|m}$. This correlation function can be related to the spectral density via
\begin{equation}
C_m(t) = \int_0^\infty d\omega J_m(\omega) [\cos(\omega t) \coth(\beta \omega / 2) - i \sin (\omega t)].
\label{eq:C_m_t}
\end{equation} We have introduced notation for the commutator $X^\times Y = XY - YX$ and the anticommutator $X^\circ Y = XY + YX$, and the short-hand notation $V_m =  \proj{m}{m}$.

A direct numerical evaluation of $\rho(t)$ according to the propagator $\tilde{\UU}(t)$ would require application of the time-ordering operator $\TTp$. This can be avoided by introduction of auxiliary operators, resulting in a system of differential equations which can be solved by standard numerical approaches. For the correlation functions which can be written as a sum of exponential contributions, $C_{m}(t) = \sum_k c_{mk} e^{- \nu_k t}$, the system of differential equations has a hierarchical structure. The system of differential equations that we obtain is in principle infinite, and thus must be truncated in a numerical evaluation. Truncating the hierarchy at an appropriate level, and treating only $k\leq K$ terms in the decomposition $C_m(t)$ exactly, and others with the Markovian approximation\cite{Ishizaki2005}, we obtain the following system of HEOM equations,
\begin{equation}
\begin{aligned}
\frac{d \rho_{\bm n}(t)}{dt} & =  -i \sum_{m=1}^{N}\sum_{k=0}^{K}
\sqrt{\frac{n_{mk}}{|c_{mk}|}}\left(c_{mk}' V_m^\times  + ic''_{mk} V_m^\circ \right)
\rho_{\bm n_{mk}^{-}}(t) \\
 + & \!\! \left(\! \LL_{\rm el} \! - \! \sum_{m=1}^{N}\sum_{k=0}^{K} \nu_{k} n_{mk} \! -  \!\! \sum_{m=1}^{N}\sum_{k=K+1}^\infty \frac{c'_{mk}}{\nu_{mk}} V_m^\times V_m^\times\right)\rho_{\bm n} (t) \\
 -  & i \sum_{m=1}^{N}\sum_{k=0}^{K} \sqrt{(n_{mk}+1)|c_{mk}|} V_m^\times \rho_{\bm n_{mk}^{+}} (t).
\end{aligned}
\label{eq:heom_scaled_with_matsubara}
\end{equation}
Auxiliary operators $\rho_{\bm n}$ are indexed by a matrix $\bm n$ with integer entries $n_{mk}$, where $m \in [1,N]$ and $k \in [0, K]$. Matrix indexes denoted by $\bm n_{mk}^{+}$ and $\bm n_{mk}^{-}$ correspond to the matrix index $\bm n$ with element $n_{mk}$ increased/decreased by one.
The electronic density matrix $\rho(t)$ in the above system of differential equations is equal to $\rho_{\bm n = 0}(t)$, i.e., with operator indexed by the zero matrix.

The level at which the hierarchy of equations is truncated can be predetermined by specifying a fixed hierarchy depth~\cite{Ishizaki2009} $N_{\rm max}$, in which case the total number of auxiliary operators is given by $N_{\rho} = ([N(K+1) + N_{\rm max}]!)/([N(K+1)]!N_{\rm max}!)$. Hierarchy depth $N_{\rm max}$ must be chosen large enough to obtain converged result. The number of auxiliary operators $N_{\rho}$ however grows very fast with increasing number of pigments $N$ and number of exponential terms $K$. Alternatively, the truncation level can be chosen adaptively\cite{Shi2009} in the process of solving the system of HEOM equations, based on the norm of the auxiliary operators. Such truncation scheme reduces number of required auxiliary operators significantly and was thus also employed in our analysis.

The exact time-dependent kernel $\KK(t)$ as well as the corresponding Markovian kernel $\mk \KK$ can be obtained from the above system of differential equations \eqref{eq:heom_scaled_with_matsubara} with the help of the Nakajima-Zwanzig formalism. The system of equations \eqref{eq:heom_scaled_with_matsubara} can be formally written as $\frac{d {\bm \rho(t)}}{dt} = \mathcal{A} {\bm \rho}(t)$, where $\bm{\rho}$ is a vector of all auxiliary operators, $\bm \rho = (\rho_{\bm n})$, and $\mathcal{A}$ is the HEOM operator defined by Eq.~\eqref{eq:heom_scaled_with_matsubara}. Introducing projection operator to the electronic density matrix as $\PP_H ( \rho, \rho_{\bm n_1}, \rho_{\bm n_2}, \dots ) = (\rho, 0, 0, \dots)$, the exact time-dependent kernel can be written as
\begin{equation}
\KK(t) = \PP_H \mathcal{A} \PP_H \delta(t) + \PP_H \mathcal{A} \GG_H(t) \QQ_H \mathcal{A} \PP_H
\label{eq:KK_heom}
\end{equation}
with the projector $\QQ_H = 1 - \PP_H$ and the propagator $\GG_H(t) = \exp[\QQ_H \mathcal{A} t]$. Taking into account properties of projection operators, relation \eqref{eq:KK_heom} enables efficient calculation of $\KK(t)$.

The method can be further optimized for the calculation of the Markovian kernel $\mk \KK$ by formal time-integration of Eq.~\eqref{eq:KK_heom}, resulting in
\begin{equation}
\mk \KK = \PP_H \mathcal{A} \PP_0 \PP_H,
\label{eq:mkKK_proj}
\end{equation}
where $\PP_0$ is a projector to the null-space of the $\QQ_H \mathcal{A}$ defined as $\PP_0=\sum_i \pproj{\PP^R_i}{\PP^L_i}$,
with $\QQ_H \mathcal{A} \kket{\PP^R_i} = 0$, $\bbra{\PP^L_i}\QQ_H \mathcal{A} = 0$, and $\bbrakket{\PP^L_i|\PP^R_j}=\delta_{ij}$, where $\delta_{ij}$ is the Kronecker delta. All $N^2$ left zero-eigenvectors $\bbra{\PP^L_i}$ can be determined directly from $\bbra{\PP^L_i} \QQ_H = 0$. Thus to evaluate $\mk \KK$, only $N^2$ right zero-eigenvectors $\kket{\PP^R_j}$ must be determined numerically.

\subsection{F\"orster theory}

The F\"orster theory is applicable when the exciton coupling between pigments is smaller that the coupling with the environment, namely, $V_{mn} \ll \lambda$ in Eqs. \eqref{eq:H_el} and \eqref{eq:H_el-ph}. Appropriate partitioning of the Hamiltonian $H$ for the perturbative treatment is thus $H_0 = \sum_m E_m \proj{m}{m} + H_{\rm ph} + H_{\rm el-ph}$ and $H_I = \frac{1}{2}\sum_{mn} V_{mn} \proj{m}{n}$. Employing Eq.~\eqref{eq:mkK_second_order_explicit} we obtain\cite{Yang2002}
\begin{equation}
\mk K_{mn}= 2 |V_{mn}|^2 \RE \int_0^\infty dt \, \mathscr{A}_m(t) \mathscr{F}_n^*(t) .
\end{equation}
where $\mathscr{A}_m(t) = e^{- i E_m t} e^{-g_m(t)}$ and $\mathscr{F}_n(t) = e^{-i(E_n - 2 \lambda_n)t}e^{-g_n(t)^*}$ are related to the absorption and fluorescence spectra of individual pigment molecules. The function $g_m(t) =  \int_0^t dt_1 \int_0^{t_1} dt_2 \, C_m(t_2)$ is known as a \textit{line broadening function}, because it determines the shape of the absorption/fluorescence spectral line via $\mathscr{A}_m(t)$ and $\mathscr{F}_m(t)$.

\subsection{Redfield theory}

The Redfield theory is applicable when the coupling of electronic and vibrational DOFs is small. The Hamiltonian $H$ is thus partitioned to $H_0 = H_{\rm el} + H_{\rm ph}$ and $H_I = H_{\rm el-ph}$. For the Redfield theory, the quantum kernel $\mk \KK$ can be evaluated according to Eq.~\eqref{eq:mkKK_second_order}. We shall evaluate it in the \textit{exciton} basis, spanned by eigenvectors of the electronic Hamiltonian $H_{\rm el} \ket{\alpha} = E_\alpha \ket{\alpha}$. In this basis we obtain
\begin{equation}
\begin{aligned}
\mk \KK_{\mu\nu,\mu'\nu'} = &  -i \delta_{\mu\mu'}\delta_{\nu\nu'} \omega_{\mu\nu} +
\varGamma_{\nu' \nu,\mu \mu'}(\omega_{
  \nu'\mu}) +
\varGamma^*_{\mu' \mu,\nu\nu'}(\omega_{\mu' \nu})  \\
 & - \delta_{\nu\nu'} \sum_\kappa \varGamma_{\mu \kappa, \kappa \mu'}(\omega_{\nu' \kappa})
 - \delta_{\mu\mu'} \sum_\kappa \varGamma^*_{\nu \kappa,\kappa \nu'}(\omega_{\mu' \kappa}),
\end{aligned}
\label{eq:mkKK_redfield}
\end{equation}
where we have introduced $\varGamma_{\mu\nu,\mu'\nu'}(\omega)= \sum_{m} \braket{\mu|V_m|\nu}\!\braket{\mu'|V_m|\nu'}\tilde C_{m}(\omega),$
and $\tilde C_m(\omega) = \int_0^\infty dt \, e^{i \omega t} C_m(t)$. From the quantum kernel $\mk \KK$, the corresponding rate kernel $\mk K$ in an arbitrary basis can be obtained using the Nakajima-Zwanzig formalism with the projection operator $\PP_p$ from Eq.~\eqref{eq:PPp}. Note that the Redfield rate kernel $\mk K$ in the exciton basis can be also obtained directly from Eq.~\eqref{eq:mkK_second_order_explicit}, resulting in rates\cite{Yang2002} $\mk K_{\alpha\beta} = \mk \KK_{\alpha \alpha,\beta\beta}$.

\subsection{Modified Redfield theory}

The modified Redfield theory extends the validity of the perturbative treatment to the range of parameters where the coupling of electronic and vibrational DOFs is not small, by including a part of $H_{\rm el-ph}$ interaction into the exactly solvable $H_0$ via a prescription $H_0 = H_{\rm el} + H_{\rm ph} + \sum_{\alpha} \ket{\alpha}\!\braket{\alpha| H_{\rm el-ph}| \alpha}\!\bra{\alpha}$, with the remaining perturbative part $
H_I = \sum_{\alpha\neq\beta} \proj{\alpha}{\alpha} H_{\rm el-ph} \proj{\beta}{\beta}
$.
The basis $\{ \ket{\alpha} \}$ is the exciton basis $H_{\rm el} \ket{\alpha} = E_\alpha \ket{\alpha}$. In this basis the rate kernel can be evaluated according to Eq.~\eqref{eq:mkK_second_order_explicit} as\cite{Yang2002}
\begin{equation}
\mk K_{\alpha\beta}= 2 \RE \int_0^\infty dt \, \mathscr{\tilde A}_\alpha(t) \mathscr{
\tilde F^*}_\beta(t) \mathscr{\tilde N}_{\alpha \beta}(t)
\end{equation}
where we have introduced functions
\begin{align}
\mathscr{\tilde A}_\alpha(t) & =  e^{-i E_{\alpha} t} e^{-g_{\alpha\alpha\alpha\alpha}(t)} \\
\mathscr{\tilde F}_\beta(t) & = e^{-i(E_\beta - 2 \lambda_{\beta\beta\beta\beta}) t} e^{-g^*_{\beta\beta\beta\beta}(t)} \\
\begin{split}
\mathscr{\tilde N}_{\alpha \beta}(t) & =
\left[\ddot g_{\beta\alpha\alpha\beta}(t) - \{ 2i \lambda_{\beta\beta\alpha\beta}-\dot g_{\alpha\alpha\alpha\beta}(t)+\dot g_{\beta\beta\alpha\beta}(t) \} \right. \\
 & \left. \{ 2i \lambda_{\beta\beta\beta\alpha} - \dot g_{\alpha\alpha\beta\alpha}(t) + \dot g_{\beta\beta\beta\alpha}(t) \}
  \right] e^{2(i \lambda_{\beta\beta\alpha\alpha} t + g_{\alpha\alpha\beta\beta}(t))}
\end{split}
\end{align}
with
$ g_{\alpha\beta\alpha'\beta'}(t)=\sum_{m=1}^N a_{\alpha\beta}^m a_{\alpha'\beta'}^m g_m(t) $ and $
\lambda_{\alpha\beta\alpha'\beta'}=\sum_{m=1}^N a_{\alpha\beta}^m a_{\alpha'\beta'}^m \lambda_m.$ The function $g_m(t)$ is the line broadening function introduced above, while $\dot g_m(t)$ and $\ddot g_m(t)$ are its first and second time-derivative. Variables $a_{\alpha \beta}^m = \braket{\alpha|m}\!\!\braket{m|\beta}$ characterize the overlap between the exciton and the site basis states.

\subsection{Variational master equation}

The variational master equation is supposed to extend the validity of perturbative approach by first transforming the Hamiltonian $H \rightarrow \tilde H$, and then fixing the exactly solvable part $\tilde H_0$ and the perturbative part $\tilde H_I$. Choosing the transformation appropriately, the perturbative part $\tilde H_I$ might be kept small over a wide range of parameters. In the context of EET \textit{variatonal polaron transformation} is usually employed, which is defined via relation $\tilde H = e^G H e^{-G}$, with $G = \sum_{m,\xi}\proj{m}{m}\omega_{m\xi}^{-1}(f_{m\xi}b_{m\xi}^\dagger - f_{m\xi} b_{m\xi})$, where $f_{m \xi}$ are free parameters of the transformation (i.e., displacement parameters). The transformed Hamiltonian is then decomposed to the exactly solvable part $\tilde H_0 = \tilde H_{\rm el} + \tilde H_{\rm ph}$ and perturbative part $\tilde H_I = \tilde H_L + \tilde H_D$.\cite{Pollock2013} Transformed electronic Hamiltonian is given by
\begin{equation}
  \tilde H_{\rm el} = \sum_{m=1}^N (E_m + R_m)\proj{m}{m} + \frac{1}{2} \sum_{m,n = 1}^N \bar{B}_m\bar{B}_n V_{mn} \proj{m}{n},
\end{equation}
while the phonon part is unchanged, $\tilde H_{\rm ph} = H_{\rm ph}$.
The perturbative part consists of $\tilde H_L$ and $\tilde H_D$ where the former is linearly dependent on bosonic operators $b_{m\xi}^\dagger$ and $b_{m \xi}$,
\begin{align}
  \label{eq:H_L}
  \tilde H_{L} = \sum_{m=1}^N \sum_{\xi} \proj{m}{m}\left[(g_{m\xi}-f_{m\xi})(b_{m\xi}^\dagger + b_{m\xi})\right],
\end{align}
while the latter is exponentially dependent on the bosonic operators,
\begin{equation}
  \label{eq:H_D}
  \tilde H_{D} = \frac{1}{2} \sum_{m,n=1}^N V_{mn}\proj{m}{n} B_{mn}.
\end{equation}
In the above expressions we have introduced the operator $B_{mn} = B_m B_n^\dagger - \bar B_m \bar B_n,$ with
$
B_m =
\exp
  \left[
    \sum_\xi \omega_{m\xi}^{-1} f_{m\xi}(b_{m\xi}^\dagger - b_{m\xi})
  \right]
$
and equilibrium expectation value $\bar B_m = \trb\{ B_m \rho_{\rm ph} \} = \exp\left[ -\frac{1}{2}\sum_\xi \frac{f_{m\xi}^2}{\omega_{m\xi}^2}\coth\frac{\beta \omega_{m\xi}}{2} \right]$,
where the equilibrium density matrix is given by $\rho_{\rm ph} = \exp(-\beta \tilde H_{\rm ph})/Z$. The shift of site energies $R_m$ is expressed as
\begin{align}
  R_m &= \sum_{\xi} \omega_{m\xi}^{-1}\left[f_{m\xi}^2 - 2 f_{m \xi}g_{m\xi} \right].
\end{align}

For the validity of the perturbative treatment, parameters $f_{m\xi}$ must be chosen such that the contribution of $\tilde H_I$ to the dynamics of the system will be small. A standard approach is to minimize the contribution of $\tilde H_I$ to the free energy of the system, given by $A = -\beta^{-1} \ln(\tr \exp(-\beta \tilde H))$. To achieve this, we write the free energy as a sum of contributions due to the exactly solvable and the perturbative part, $A = A_0 + A_I$. We have introduced the free energy of exactly solvable part $A_0 = - \frac{1}{\beta}\ln\left[\tr\{e^{-\beta \tilde H_0}\} \right]$, while $A_I$ is the term we are trying to minimize. Using the Feynman-Bogoliubov upper bound\cite{McCutcheon2011,Zimanyi2012} we obtain inequality $A \leq A_0$. To minimize $A_I$ we must thus minimize $A_0$, leading to the minimization condition with respect to all transformation parameters,
\begin{equation}
\pd{A_0}{f_{m\xi}}=\pd{A_0}{R_m}\pd{R_m}{f_{m\xi}} + \pd{A_0}{\bar{B}_m}\pd{\bar{B}_m}{f_{m\xi}}=0.
\label{eq:minimization-discrete}
\end{equation}

For a continuous distribution of environmental oscillators $J(\omega)$ we can introduce a frequency-dependent displacement function $F_{m}(\omega)$, which determines displacement parameters via relation $f_{m\xi} = F_m(\omega_{m\xi})g_{m\xi}$, leading to the minimization condition \eqref{eq:minimization-discrete} in the form
\begin{equation}
\label{eq:Fm_omega}
F_m(\omega ,\{R_m,\bar{B}_m\})=\frac{2 \omega \pd{A_0}{R_m}}{2\omega \pd{A_0}{R_m}-\bar{B}_m \pd{A_0}{\bar{B}_m} \coth(\beta \omega /2)},
\end{equation}
which defines a set of $2N$ coupled integral equations, which must be solved self-consistently for the unknown $F_m(\omega)$.

The approximate quantum kernel $\tilde{\mk \KK}$ for the perturbative treatment of $\tilde H_I$ is obtained from Eq.~\eqref{eq:mkKK_second_order}, resulting in a somewhat lengthy expression as different combinations of perturbative Hamiltonians $H_L$ and $H_D$ must be accounted for in the expansion. The final expression in the site basis $\tilde{\mk \KK}_{mn,m'n'}$ is given in Appendix~\ref{app:variational}, Eq.~\eqref{eq:mkKK_variational}. Note that the final expression for the kernel is not identical as the one in Ref.~\onlinecite{Pollock2013}, where the resulting equations are obtained employing the Markovian approximation in interaction picture, thus describing short-time dynamics of the electronic density matrix.

The kernel $\tilde{\mk \KK}$ determines stationary states and currents in the transformed frame. To obtain description in the original site basis, we have to transform it back to the original frame. The inverse polaron transformation in the system subspace $ \tilde  \rho \rightarrow \rho$ depends on the state of the whole system $\tilde R$, which was however projected out. We thus have to assume the form of $\tilde R$. If the variational polaron transformation is indeed such that $\tilde H_I$ is small, $\tilde R$ can be approximated as $\tilde R \approx \tilde \rho \otimes \rho_{\rm ph}$. In such case, the inverse polaron transformation for the density matrix is given by $\rho_{mn}(t) = \left([1-\delta_{mn}]\bar{B}_m\bar{B}_n + \delta_{mn}\right) \tilde \rho_{mn}(t)$, and thus the kernel in the original frame is related to the kernel in the transformed frame by
\begin{equation}
\mk \KK_{mn,m'n'} = \frac{[1-\delta_{mn}]\bar{B}_m\bar{B}_n + \delta_{mn}}{[1-\delta_{m'n'}]\bar{B}_{m'}\bar{B}_{n'} + \delta_{m'n'}} \tilde{\mk \KK}_{mn,m'n'}.
\label{eq:KK_trans_back}
\end{equation}

\section{Comparison}
\label{sec:comparison}

In this section we employ the methods presented above to evaluate rate kernels for two simple PPCs, consisting of two (dimer) and three (trimer) pigment molecules. When the underlying method provides us with the quantum kernel $\mk \KK$, the corresponding rates in the site basis $\mk K_{mn}$ and the exciton basis $\mk K_{\alpha\beta}$ are evaluated using the Nakajima-Zwanzig formalism with the projection operator $\PP_p$, Eq.~\eqref{eq:PPp}. This is possible for the Redfield theory and the variational master equation. From the F\"orster theory however, we can only obtain rates in the site basis $\mk K_{mn}$, and from the modified Redfield theory only in the exciton basis $\mk K_{\alpha\beta}$. For the comparison of approximate rates with the exact rates obtained by the HEOM method, we observe a ratio between them,
\begin{equation}
\sigma_{mn} = \frac{\mk K_{mn}}{\mk K_{mn}^{\rm exact}}.
\label{eq:sigma_mn}
\end{equation}
The expression in the exciton basis $\sigma_{\alpha\beta}$ is defined analogously.

\subsection{Dimer}

The electronic Hamiltonian $H_{\rm el}$ for a dimer system is uniquely determined by two parameters, the site energy $E = E_2 - E_1$, and the exciton interaction between the two pigments $V=V_{12}$. For simplicity, each site is coupled to an independent phonon bath with a Drude-Lorentz spectral density \eqref{eq:drude-lorentz} with the same value of the reorganization energy $\lambda$ and the cutoff frequency $\gamma$. We evaluate rate kernels for a range of reorganization energies $\lambda$ and site energies $E$, for weakly ($V=20\, {\rm cm}^{-1}$) and strongly ($V=100\,{\rm cm}^{-1}$) coupled pigment molecules, and all at fixed bath relaxation time $\gamma^{-1} = 166 \, {\rm fs}$ and temperature $T=300 \, {\rm K}$.

\begin{figure}
    \includegraphics[width=1\columnwidth]{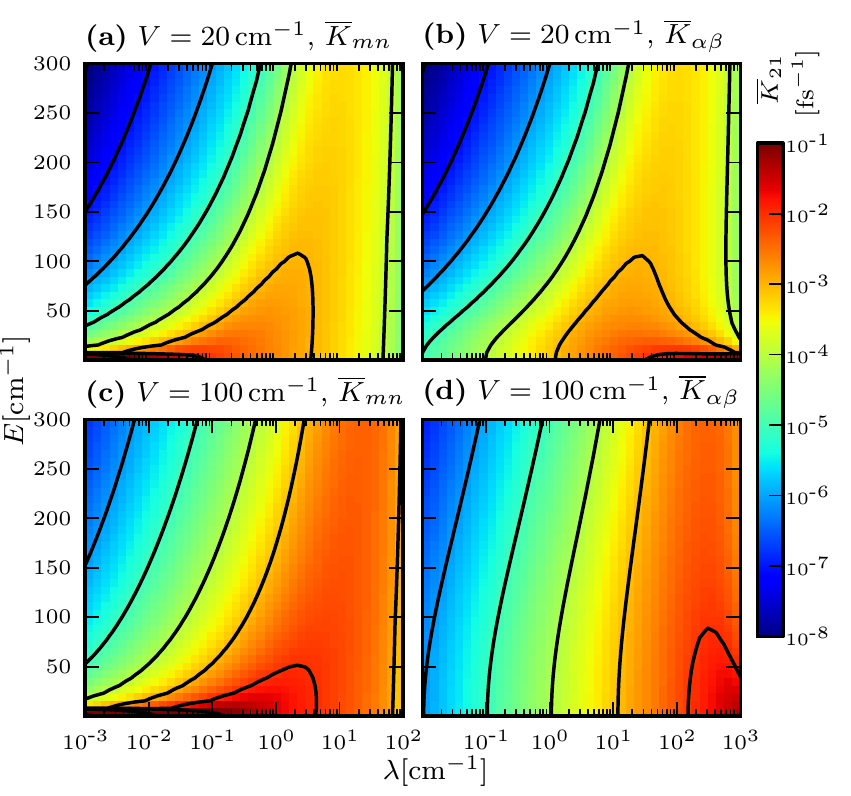}
    \caption{\label{fig:plot_heom_rates}
    Exact rates $\mk K_{21}$ for dimer in site basis (a),(c) and exciton basis (b),(d), at two values of exciton coupling $V$. The rates were obtained using the HEOM method. Other parameters are $T=300 \, {\rm K}$, $\gamma^{-1} = 166\,{\rm fs}$.
    }
\end{figure}

Exact rates were obtained using the HEOM method in both site and exciton basis, $\mk K_{mn}^{\rm exact}$ and $\mk K_{\alpha\beta}^{\rm exact}$. Results are shown in Fig.~\ref{fig:plot_heom_rates}. Parameters for the HEOM method (i.e., truncation depth of the hierarchy and the number of exactly treated correlation function expansion terms $K$) were chosen such that the relative error of the obtained rates is less than $10^{-4}$ within the whole parameter space considered. Differences in the overall behavior of rates in the site and the exciton basis as $\lambda \rightarrow 0$ and $E \rightarrow 0$ can be understood by considering unitary dynamics of isolated electronic part, governed by $H_{\rm el}$. Two electronic eigenstates $\ket{\alpha}$ of the dimer system at $E = 0$ have the same overlap with both site basis states $\ket{m}$. Thus, choosing the initial state as a singly-occupied site, $\rho(0)=\proj{m}{m}$, results in oscillations of the population of the other site, leading to large transfer rates in the site basis. If one instead takes a singly-occupied exciton as an initial state, $\rho(0)=\proj{\alpha}{\alpha}$, population of the other exciton does not change with time, leading to vanishing transfer rates in the exciton basis.

\begin{figure}
  \begin{minipage}{1\linewidth}
    \label{fig:plot_rate_comp_V20}
    \includegraphics[width=1\columnwidth]{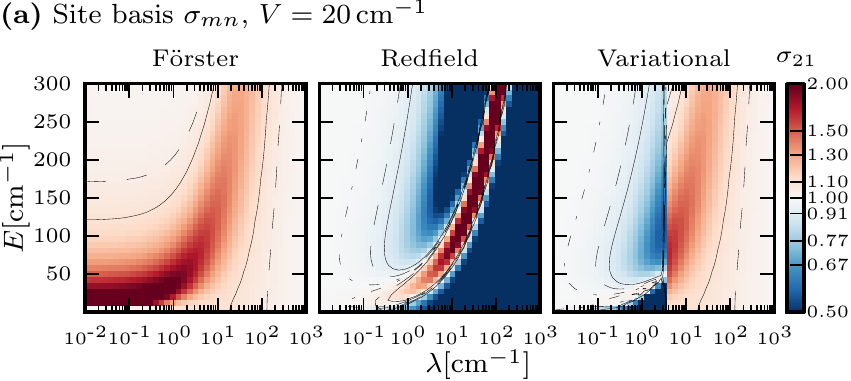}
  \end{minipage}
  \begin{minipage}{1\linewidth}
    \label{fig:plot_rate_comp_V100}
    \includegraphics[width=1\columnwidth]{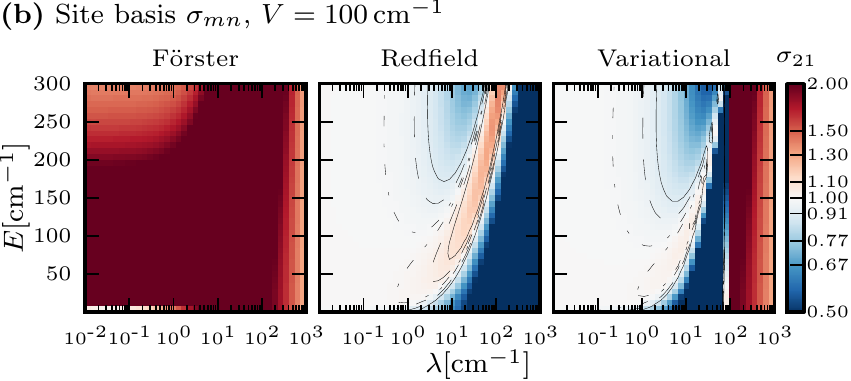}
  \end{minipage}
  \caption{\label{fig:plot_rate_comp}
    Ratios between approximate and exact rates in the site basis $\sigma_{21}$ for a dimer, obtained for a range of $\lambda$ and $E$ at two exciton coupling strengths $V$. Solid contours correspond to $\sigma=(1.1^{-1},1.1)$, dashed contours to $\sigma=(1.05^{-1},1.05)$ and dash-dotted contours to $\sigma=(1.01^{-1},1.01)$. Other parameters are $T=300 \, {\rm K}$, $\gamma^{-1} = 166\,{\rm fs}$.
  }
\end{figure}

\begin{figure}
  \begin{minipage}{1\linewidth}
    \label{fig:plot_rate_comp_V20_E}
    \includegraphics[width=1\columnwidth]{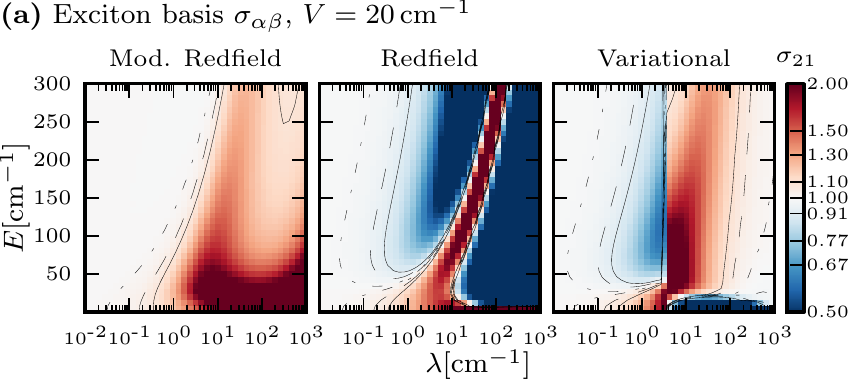}
  \end{minipage}
  \begin{minipage}{1\linewidth}
    \label{fig:plot_rate_comp_V100_E}
    \includegraphics[width=1\columnwidth]{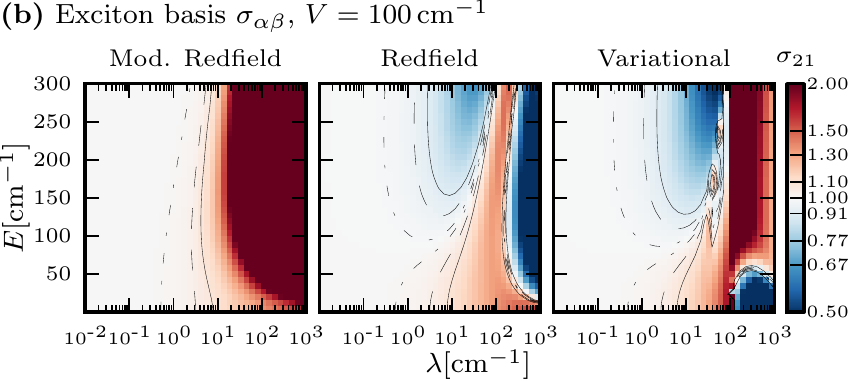}
  \end{minipage}
  \caption{
    \label{fig:plot_rate_comp_E}
    Ratios between approximate and exact rates in the exciton basis $\sigma_{21}$ for a dimer, obtained for a range of $\lambda$ and $E$ at two exciton coupling strengths $V$. Solid contours correspond to $\sigma=(1.1^{-1},1.1)$, dashed contours to $\sigma=(1.05^{-1},1.05)$ and dash-dotted contours to $\sigma=(1.01^{-1},1.01)$. Other parameters are $T=300 \, {\rm K}$, $\gamma^{-1} = 166\,{\rm fs}$.
    }
\end{figure}

\begin{figure}
    \includegraphics[width=1\columnwidth]{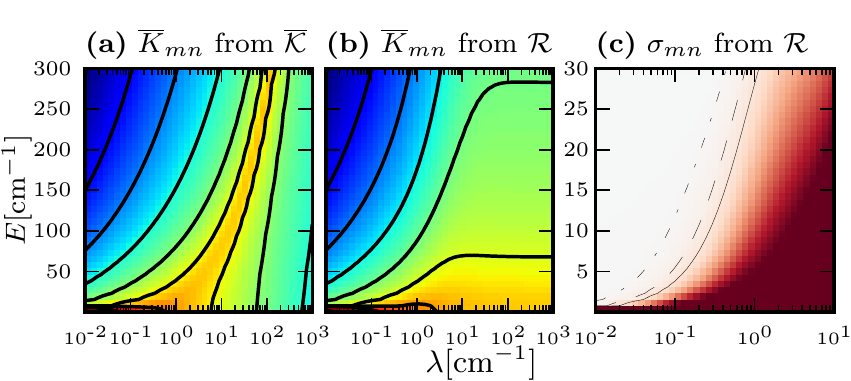}
    \caption{\label{fig:plot_rate_red}
    Comparison of rate $\sigma_{21}$ obtained from two different versions quantum kernel for the Redfield theory, (a) the one obtained in this work $\mk \KK$, Eq.~\eqref{eq:mkKK_redfield}, and (b) the standard Redfield tensor $\mathcal{R}$, Eq.~\eqref{eq:redfield_tensor}. Fig.~(c) is showing a relative error in the site basis rates $\sigma_{mn}$ obtained from the Redfield tensor $\mathcal{R}$. For $E\rightarrow 0$, invalid rates are obtained even for small environmental coupling $\lambda \rightarrow 0$. Parameters are the same as in Fig.~\ref{fig:plot_rate_comp}, with interaction strength $V=20\,{\rm cm}^{-1}$.
    }
\end{figure}

In Fig.~\ref{fig:plot_rate_comp} ratios in the site basis $\sigma_{mn}$ between  approximate and exact rates are shown for the F\"orster method, the Redfield method and the variational master equation. In Fig.~\ref{fig:plot_rate_comp_E} ratios in the exciton basis $\sigma_{\alpha\beta}$ are shown for the modified Redfield method, the Redfield method and the variational master equation. Colors in the plots signify how well each approximate method reproduces the exact rates. Color coding is chosen such that regions with blue color correspond to approximate rates that are less than half of the exact rate, while regions with red color denote approximate rates that are more that twice as large as the exact rate, while intermediate colors denote ratios in-between. White color denotes regions where approximate method matches the exact rate. Contour lines provide detailed information on the accuracy of approximate methods, denoting regions where approximate rate is within $10 \%$ (solid line), $5\%$ (dashed line) or $1 \%$ (dash-dotted line) of the exact value.

F\"orster rates match the exact rates reasonably well if $\lambda \gg V$ or $E \gg V$, which is what is expected from the separation of the exact and perturbative part in the derivation. For weak pigment coupling $V = 20 \, {\rm cm}^{-1}$, in the parameter ranges of real PPCs ($\lambda \gtrsim 10 \, {\rm cm}^{-1}$), the F\"orster theory results in rates that are within $\sim 20\%$ of the exact rates. Increasing pigment coupling strength $V$, the validity of the F\"orster theory moves out of the range of relevant reorganization energies $\lambda$, e.g., at $V=100\, {\rm cm}^{-1}$, rates obtained at $\lambda \sim 100 \, {\rm cm}^{-1}$ are off by a factor of 2. Nonetheless the overall dependence of the F\"orster rate $\mk K_{21}$ on the parameters $\lambda$ and $E$ is the expected one - it reaches maximal value going from small to large $\lambda$, and decreases with increasing site energy difference $E$.

The Redfield rates are expected to be valid for small reorganization energies $\lambda \ll V$. For the parameters considered, this corresponds to quite small reorganization energies (smaller than one can realistically expect in PPCs). E.g., for a weak pigment coupling $V=20 \, {\rm cm}^{-1}$, the rates are within $20 \%$ of the exact values for $\lambda \lesssim 1\,{\rm cm}^{-1}$, and for a strong pigment coupling $V=100 \, {\rm cm}^{-1}$ for $\lambda \lesssim 10 \,{\rm cm}^{-1}$. Within this limited range range of validity, the Redfield theory results in correct rates in both, the site and the exciton basis. Also, the overall dependence of rates on parameters $\lambda$ and $E$ is again as expected (see Fig.~\ref{fig:plot_rate_red}, first plot). The results in Fig.~\ref{fig:plot_rate_red} however do not comply with the analysis for the Redfield equation in Ref.~\onlinecite{Ishizaki2009a} (see, e.g., Fig.~2 in the cited reference), where the Redfield rate has a plateau at large reorganization energies $\lambda$. Such behavior at large $\lambda$ is indeed observed when rates are obtained from the standard Redfield tensor, given by~\cite{Ishizaki2009a,breuer2002theory,may2011charge}
\begin{equation}
\begin{split}
\mathcal{R}_{\mu\nu,\mu'\nu'} = & -i \delta_{\mu\mu'}\delta_{\nu\nu'} \omega_{\mu\nu} + \varGamma_{\nu'\nu\mu\mu'}(\omega_{\mu'\mu}) + \varGamma_{\mu'\mu\nu\nu'}^*(\omega_{\nu'\nu}) \\
& - \delta_{\nu\nu'}\sum_\kappa \varGamma_{\mu\kappa\kappa\mu'}(\omega_{\mu'\kappa})
- \delta_{\mu\mu'}\sum_\kappa \varGamma_{\nu\kappa\kappa\nu'}^*(\omega_{\nu'\kappa}).
\end{split}
\label{eq:redfield_tensor}
\end{equation}
The rates corresponding to $\mathcal{R}$ are shown in Fig.~\ref{fig:plot_rate_red}b. The observed behavior for $\mathcal{R}$ is due to the Markovian approximation being done in the \textit{interaction} picture, which results in a correct short-time dynamics of the density matrix $\rho(t)$, however, it does not result in the appropriate steady-state rates. Even for small $\lambda$, where the perturbative Redfield approach is expected to work well, the site-basis rates $\mk K_{mn}$ obtained from the standard Redfield tensor $\mathcal{R}$, Eq.~\eqref{eq:redfield_tensor}, do not match the exact rates. This is demonstrated in Fig.~\ref{fig:plot_rate_red}c, where a ratio between rates in the site basis obtained from $\mathcal{R}$ and exact rates is shown in a region of small $\lambda$ and $E$. As the site energy difference $E$ approaches zero, the standard Redfield tensor results in wrong site-basis rates independently of the reorganization energy $\lambda$. Note, however, that the rates in the exciton basis $\mk K_{\alpha\beta}$ obtained from the Redfield tensor $\mathcal{R}$ comply with the exact rates at $\lambda \sim 0$ for an arbitrary site energy $E$.

Whenever the site energy $E \gg V$, the modified Redfield theory extends the range of validity of the Redfield equation into the region of large reorganization energies $\lambda$ (see Fig,~\ref{fig:plot_rate_comp_E}). In such case the exciton basis states $\ket{\alpha}$ are well approximated by the site basis states $\ket{m}$, leading to small values of $a_{\alpha \beta}^m$ in the corresponding perturbative Hamiltonian $H_I$. At small pigment coupling $V=20\,{\rm cm^{-1}}$ improvement in the accuracy of obtained rates is evident for $E \gtrsim 150\,{\rm cm}^{-1}$, where rates are within $20 \%$ of the exact value for arbitrary reorganization energies $\lambda$. However, increasing pigment coupling to $V=100\,{\rm cm}^{-1}$ and for large $\lambda$, the modified Redfield equation already fails to provide significant improvement compared to the ordinary Redfield theory. Also, the modified Redfield equation only enables calculation of rates in the exciton basis.

Variational master equation was devised to extend the validity of perturbative treatments into the range of arbitrary reorganization energies. However, for the Drude-Lorentz spectral density, it merely combines together the rates obtained by the Redfield method at small $\lambda$ and the rates obtained by the F\"orster method at large $\lambda$, with a discontinuous jump at some intermediate value $\lambda_c$. Analogous behavior was observed in Refs.~\onlinecite{McCutcheon2011,Zimanyi2012}. For $\lambda > \lambda_c$, the minimization condition \eqref{eq:Fm_omega} results in a displacement function $F(\omega) = 1$. Variational master equation in this regime is equivalent to the \textit{polaron} master equation, which in fact exactly reproduces rates of the F\"orster theory for the Ohmic spectral densities.\cite{McCutcheon2011}

\subsection{Trimer}

In this section we extend the analysis of the rate kernels to a trimer system in order to identify possible nontrivial effects due to multiple sites taking part in the EET process. To uniquely specify an arbitrary trimer electronic Hamiltonian $H_{\rm el}$, 6 parameters have to be chosen. Systematic inspection of the whole parameter space would thus be cumbersome and not very insightful. Therefore, we limit ourself to $H_{\rm el}$ of the form
\begin{equation}
H_{\rm el}= \begin{pmatrix}
  240. & 87.7 & 5.5 \\
  & E_{2} &  30.8 & \\
  & & 0. \\
\end{pmatrix},
\label{eq:hamiltonian_trimer}
\end{equation}
which is the FMO Hamiltonian from Ref.~\onlinecite{Adolphs2006} for sites 1, 2 and 3 with a variable site energy $E_2$. Each site is coupled to an independent phonon bath with Drude-Lorentz spectral density \eqref{eq:drude-lorentz} with $\gamma^{-1} = 166\,{\rm fs}$.

\begin{figure}
    \includegraphics[width=1\columnwidth]{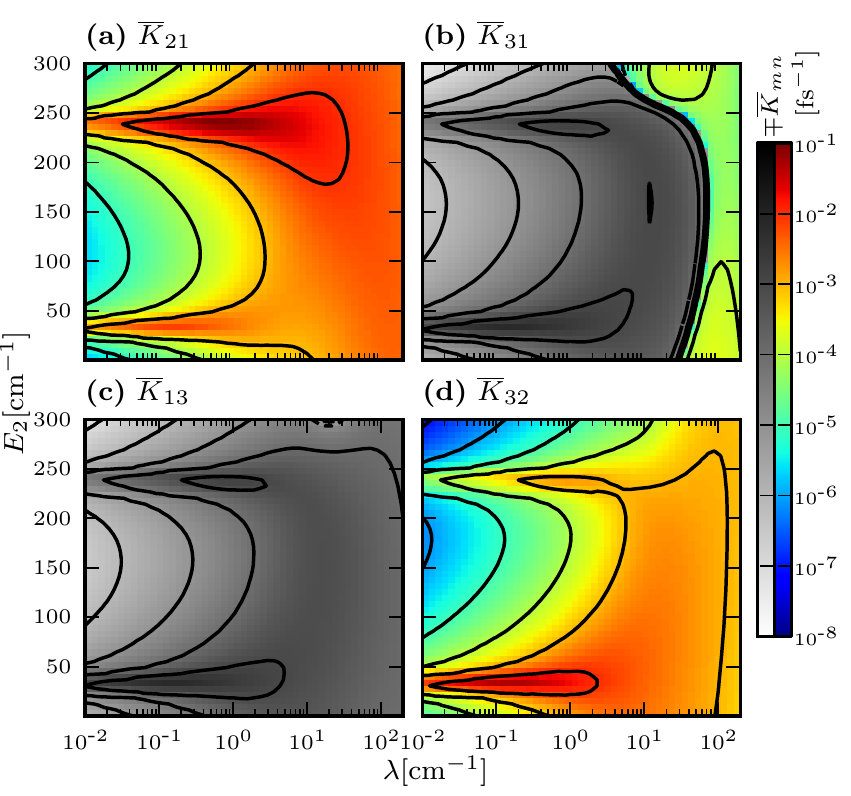}
    \caption{\label{fig:plot_heom_rates_tri}
    Exact rates $\mk K_{mn}$ in the site basis for a trimer with variable site energy $E_2$ and reorganization energy $\lambda$. Blue-to-red color coding denotes positive rates, while white-to-black coding denotes negative rates. Other parameters are $T=300 \, {\rm K}$, $\gamma^{-1} = 166\,{\rm fs}$.
    }
\end{figure}

\begin{figure}
    \includegraphics[width=1\columnwidth]{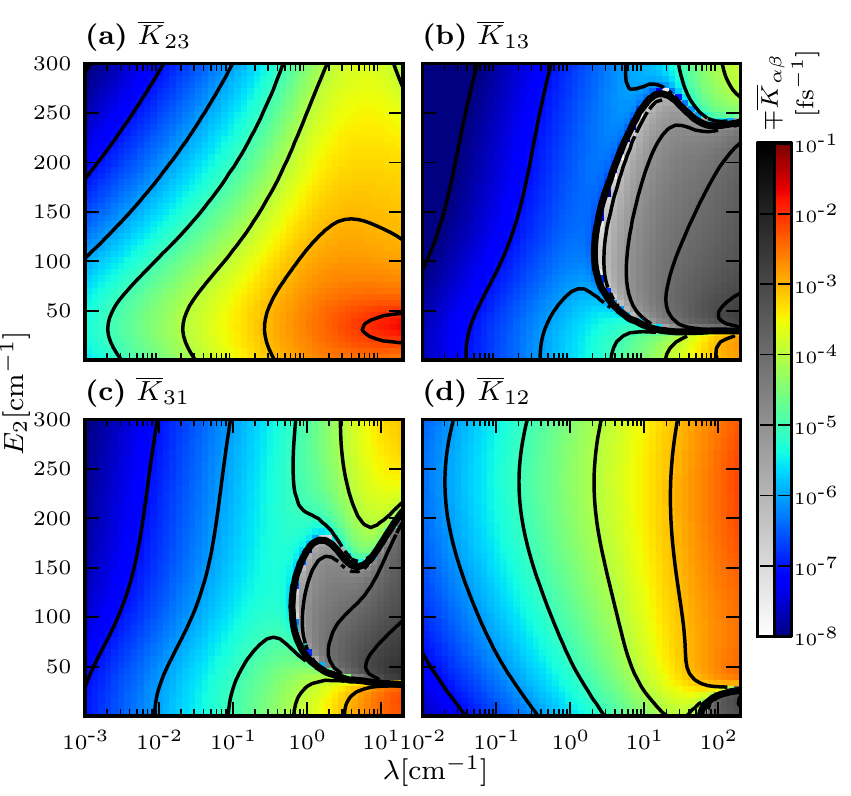}
    \caption{\label{fig:plot_heom_rates_tri_E}
    Exact rates $\mk K_{\alpha \beta}$ in the exciton basis for a trimer with variable site energy $E_2$ and reorganization energy $\lambda$. Blue-to-red color coding denotes positive rates, while white-to-black coding denotes negative rates. Other parameters are $T=300 \, {\rm K}$, $\gamma^{-1} = 166\,{\rm fs}$.
    }
\end{figure}

\begin{figure}
  \begin{minipage}{1\linewidth}
    \label{fig:plot_heom_rates_compare_tri_site}
    \includegraphics[width=1\columnwidth]{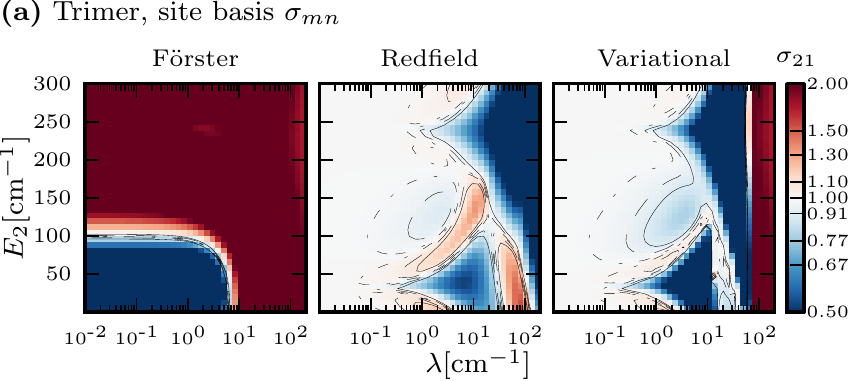}
  \end{minipage}
  \begin{minipage}{1\linewidth}
    \label{fig:plot_heom_rates_compare_tri_E}
    \includegraphics[width=1\columnwidth]{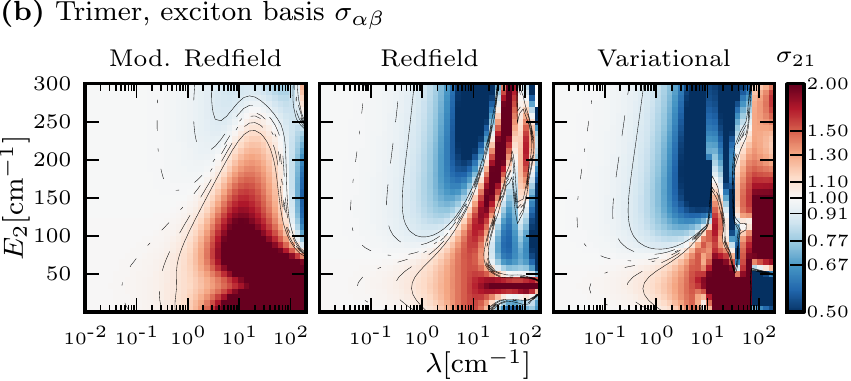}
  \end{minipage}
  \caption{
    \label{fig:plot_heom_rates_compare_tri}
    Ratio between approximate rates and exact rates $\sigma_{21}$ in (a) the site basis and (b) the exciton basis for a trimer. Rates were obtained by approximate methods for a range of $\lambda$ and $E_2$. Solid contour denotes $\sigma=(1.1^{-1},1.1)$, dashed contour $\sigma=(1.05^{-1},1.05)$ and dash-dotted contour $\sigma=(1.01^{-1},1.01)$. Rates were evaluated for parameters $T=300 \, {\rm K}$, $\gamma^{-1} = 166\,{\rm fs}$.
    }
\end{figure}

We have evaluated the exact rates for a range of values $\lambda$ and $E_2$ using the HEOM method such that the relative errors are within $\lesssim 10^{-3}$. As in the dimer case, we consider rates in the site and the exciton basis. Various rates $\mk K_{mn}$ in the site basis are shown in Fig.~\ref{fig:plot_heom_rates_tri}. Rates between most strongly coupled sites, i.e., $1 \rightarrow 2$ and $2 \rightarrow 3$, are peaked at resonant values of site energy $E_2$, while with respect to the reorganization energy $\lambda$, a maximum is again attained at intermediate $\lambda$. For a weakly coupled sites 1 and 3 however, rates between them can be negative (denoted by white-to-black color coding in plots). Negative rates prevent a simple interpretation of the classical master equation~\eqref{eq:master_eq_classical} as a hopping process between sites, where the probability of a jump is given by a product of a rate and a population of the site. For a discussion of possible interpretation of negative rates see Ref.~\onlinecite{Laine2012}. Nonetheless, the kernel $\mk K_{mn}$ determines the appropriate stationary state, which further determines the efficiency of the EET. Note that with increasing reorganization energy $\lambda$, the EET process becomes more incoherent, rendering rates $\mk K_{mn}$ positive again. For an example see second plot of Fig.~\ref{fig:plot_heom_rates_tri} showing rate $K_{31}$. The reverse rate $K_{13}$ however remains negative for the whole range of reorganization energies considered, $\lambda < 200\,{\rm cm}^{-1}$. Negative rates are also obtained in the exciton basis, shown in Fig.~\ref{fig:plot_heom_rates_tri_E}. Exciton states are numbered according to their energy, starting from the lowest energy exciton state. Negativity of rates in the exciton basis however appears at large reorganization energies -- just in the opposite regime as in the site basis. Thus, at small $\lambda$, dynamics can be interpreted as a jump process between excitons.

Comparing the exact rates to the approximate rates, Fig.~\ref{fig:plot_heom_rates_compare_tri}, we observe behavior analogous to that in the dimer system, with each perturbative method being valid in its corresponding regime. However, as we approach the limits of  validity, one can not reliably state whether the approximate method will result in the overestimated or underestimated rates. For example, while the F\"orster and the modifier Redfield method generally overestimated rates in the case of the dimer system, for the trimer system they can either overestimate or underestimate them, depending on the parameters of the model. Rates obtained by the variational master equation now display multiple discontinuous jumps as $H_{\rm el}$ and $\lambda$ is varied, due to various local minima of $A_0$ becoming global minima in different parameter regimes, resulting in a sudden change of the displacement function $F_m(\omega)$.~\cite{Lee2012,Pollock2013}

\section{Conclusion}
\label{sec:conclusion}

We have compared EET rates obtained by various perturbative approaches to the exact rates obtained by the HEOM method on the basis of a common criteria, namely the rate kernel $\mk K$. The exact HEOM method was optimized for the calculation of the time-integrated kernel, translating the problem to the evaluation of a null-space of large sparse matrix, for which efficient numerical procedures exist. Together with the adaptive truncation scheme, the method can be used for determination of exact rates for relatively large PPCs at intermediate reorganization energies, e.g., rates for the FMO complex with 7 pigment molecules can be obtained within hours on a standard PC. Numerical aspects of the HEOM method were inspected for the Drude-Lorentz spectral density, where a hierarchical structure of the HEOM operator is obtained.

The exact rates were calculated for a dimer and a trimer system. In the case of a dimer system, the obtained rates are positive both in the site and in the exciton basis, independently of the parameters of the system. Positive rates enable interpretation of the process as a classical jump-process, where each rate determines probability of a jump to the corresponding state. In the trimer system however, some rates can become negative, depending on the basis in which they are evaluated and on the parameters of the model. For example, rates in the exciton basis become negative when the reorganization energy $\lambda$ is increased, while at small $\lambda$ rates in the site basis become negative. This suggests that at intermediate values of the reorganization energy $\lambda$ there might exist a basis $\ket{M}$, interpolating between the site and the exciton basis, such that the corresponding rate kernel $\mk K_{MN}$ would result in positive rates between the basis states, enabling a jump-like interpretation of the EET process withing the whole parameter regime. Determination of the appropriate basis might be related to the notion of preferred~\cite{Balevicius2012} (or global~\cite{Gelzinis2011}) basis.

Perturbative approaches for the calculation of EET rates have the advantage of being numerically efficient, while also providing more insight into the underlying processes in the corresponding parameter regime. However, based on the analysis presented here, none of the approximate approaches considered can be reliably used for the determination of EET rates in the relevant intermediate electron-phonon coupling regime. Even more, one can not say whether a given approximate method will either overestimate or underestimate rates. This depends intricately on the parameters of the model, as seen in the trimer example.

Variational master equation, which was devised to work for an arbitrary reorganization energy $\lambda$, also results in inaccurate rates for intermediate reorganization energies $\lambda$. It could be that the variational polaron transformation is not sufficiently general to render interaction Hamiltonian $\tilde H_{I}$ small enough for perturbative treatment at arbitrary $\lambda$. Alternatively, the criteria based on free energy $A$ may not give the appropriate minimization condition, resulting in suboptimal displacement function $F(\omega)$. The underlying cause of the failure of the variational polaron transformation could be possibly identified with the help of exact calculations. With this knowledge, variational master equation could be in principle further improved to work reliably also in the region of intermediate reorganization energies $\lambda$. Such improvement would be very beneficial for the analysis of EET in larger PPCs, where exact calculations become inefficient.

\section{References}

\bibliography{rate_comp}

\appendix

\section{Quantum kernel \texorpdfstring{$\tilde{\mk \KK}$}{K} for variational master equation}
\label{app:variational}

In the following, we present the expression for the quantum kernel $\tilde{\mk \KK}$ for the variational master equation, obtained from Eq.~\eqref{eq:mkKK_second_order} with appropriate separation of exact and perturbative Hamiltonians $H_0$ and $H_I$. In the site basis $\{ \ket{m} \}$ we obtain
\begin{widetext}
\begin{align}
\label{eq:mkKK_variational}
\tilde{\mk \KK}_{mn,m'n'} =& I^0_{mn,m'n'} +
  \int_0^\infty dt \, \left[ I^{LL}_{mn,m'n'}(t) + I^{DD}_{mn,m'n'}(t) + I^{LD}_{mn,m'n'}(t) + I^{DL}_{mn,m'n'}(t) \right],\\
  I^{LL}_{mn,m'n'} (t) =& - \Uv{mm'}{n'n} \left[C_{LL}^{mm'}(t) - C_{LL}^{m'n}(-t) - C_{LL}^{nm'}(t) + C_{LL}^{n'n}(-t)  \right],
  \label{eq:I_LL}\\
I^{DD}_{mn,m'n'} (t) =& \sum_{q,p} \left[
- \Uv{qp}{n'n} C_{DD}^{mqpm'}(t)
+ \Uv{qm'}{pn} C_{DD}^{n'pmq}(-t)
+ \Uv{mq}{n'p} C_{DD}^{pnqm'}(t)
- \Uv{mm'}{qp}  C_{DD}^{n'qpn}(-t) \right],
\label{eq:I_DD} \\
I^{LD}_{mn,m'n'} (t)
=&
 \sum_{p}
\left[
- \Uv{mp}{n'n} C_{LD}^{mpm'}(t)
+ \Uv{mm'}{pn} C_{LD}^{mpn'}(-t)
+ \Uv{mp}{n'n} C_{LD}^{npm'}(t)
- \Uv{mm'}{pn} C_{LD}^{npn'}(-t) \right],
\label{eq:I_LD} \\
I^{DL}_{mn,m'n'}(t)
 =& \sum_{p} \left[
- \Uv{pm'}{n'n} C_{LD}^{m'pm}(t)
+ \Uv{pm'}{n'n} C_{LD}^{n'mp}(-t)
+ \Uv{mm'}{n'p} C_{LD}^{m'np}(t)
- \Uv{mm'}{n'p} C_{LD}^{n'pn}(-t)\right],
\label{eq:I_DL}
\end{align}
\end{widetext}
where $I^0_{mn,m'n'} =  -i \braket{m|[\tilde H_{\rm el}, \proj{m'}{n'}]|n}$ is the contribution due to unitary dynamics, while the remaining terms are resulting from different combinations of perturbative Hamiltonians $\tilde H_L$ and $\tilde H_D$. We have also introduced a short-hand notation $\Uv{mn}{pq} =\braket{m|e^{-i \tilde H_{\rm el} t}|n}\!\!\braket{p|e^{ i \tilde H_{\rm el} t}|q}$. In above expressions various correlation functions have been introduced, defined as $C_{LL}^{mn}(t) = \trb\{ \tilde H_L^m(t) \tilde H_L^n \rho_{\rm ph}\}$, $C_{DD}^{mnpq}(t) = \trb\{ \tilde H_D^{mn}(t) \tilde H_D^{pq} \rho_{\rm ph}\}$ and $C_{LD}^{mpq}(t) = \trb\{ \tilde H_L^m(t) \tilde H_D^{pq} \rho_{\rm ph}\}$, where the time argument denotes the interaction picture $\tilde H_{L(D)}^m(t) = e^{i H_{\rm ph} t} \tilde H_{L(D)}^m  e^{-i H_{\rm ph} t}$. Correlation functions can be further expressed with the spectral density $J_m(\omega)$ and the displacement function $F_m(\omega)$ as
\begin{widetext}
\begin{align}
C_{LL}^{mm}(t) =& \int_0^\infty \, d\omega J_m(\omega)[1-F_m(\omega)]^2
\left(\cos \omega t \coth\frac{\beta \omega}{2} - i \sin \omega t \right ),\\
C_{DD}^{mnpq}(t) = &
V_{mn}V_{pq} \bar{B}_m\bar{B}_n\bar{B}_p\bar{B}_q \left(e^{-\delta_{mp} \phi_{DD}^m(t) - \delta_{nq} \phi_{DD}^n(t)} + e^{\delta_{mq} \phi_{DD}^m(t) + \delta_{np} \phi_{DD}^n(t)} - 2 \right),\\
C_{LD}^{mpq}(t) = & (\delta_{mp} V_{mq} \bar{B}_{m} \bar{B}_{q}-\delta_{mq} V_{mp} \bar{B}_{m} \bar{B}_{p}) \phi_{LD}^m(t),
\end{align}
with
\begin{align}
\phi_{DD}^m(t) = & \int_0^\infty d\omega \, \frac{J_m(\omega)}{\omega^2} F_m(\omega)^2\left(\cos \omega t \coth\frac{\beta\omega}{2} - i \sin \omega t \right), \\
\phi_{LD}^m(t)=&
\int_0^\infty d \omega \frac{J_m(\omega)}{\omega} F_m(\omega)[1-F_m(\omega)]\left(\cos \omega t - i \sin \omega t \coth\frac{\beta\omega}{2} \right).
\end{align}
\end{widetext}

\section{Correlation functions}
\label{app:correlation_functions}

In this appendix, we collect some analytical expressions for correlation functions for Drude-Lorentz spectral density~\eqref{eq:drude-lorentz}, which can be obtained using, e.g., contour integration~\cite{mukamel1999principles}.
For the correlation function $C(t)$ from Eq.~\eqref{eq:C_m_t}, we obtain real and imaginary part of the correlation function $C(t) = C'(t) + i C''(t)$ as
\begin{align}
  \label{eq:Re_Ct_drude-lorentz}
  C'(t) & = \lambda \gamma \cot\left(\frac{\gamma \beta}{2}\right)e^{-\gamma|t|}+ \sum_{k=1}^\infty \frac{4 \lambda \gamma}{\beta} \frac{\nu_k}{\nu_k^2 - \gamma^2} e^{-\nu_k|t|}, \\
  C''(t) & = - \frac{t}{|t|} \lambda \gamma e^{-\gamma |t|},
  \label{eq:Im_Ct_drude-lorentz}
\end{align}
where the $k$-terms are known as Matsubara terms and $\nu_k = 2\pi k i / \beta$ are Matsubara frequencies. In the Redfield theory the half-sided Fourier transform of this correlation function is required, $\tilde C_m(\omega) = \int_0^\infty dt \, e^{i \omega t} C_m(t)$. It can be calculated by integration of the above expressions, resulting in
\begin{align}
  \label{eq:Re_C_half_om}
  \tilde C'(\omega) =& \frac{\pi}{2}J(\omega)\left(\coth\frac{\beta \omega}{2} + 1 \right),\\
  \tilde C''(\omega) =& \frac{\lambda \gamma}{\omega^2 + \gamma^2}\left(\omega \cot \frac{\beta \gamma}{2} - \gamma\right) + \frac{4 \lambda\gamma}{\beta}\sum_{k=1}^\infty \frac{\nu_k}{\nu_k^2-\gamma^2}\frac{\omega}{\omega^2 + \nu_k^2}.
  \label{eq:Im_C_half_om}
\end{align}

\end{document}